\documentclass[twocolumn,showpacs,preprintnumbers,amsmath,amssymb,footinbib,prl,aps]{revtex4-1}

\usepackage{graphicx}
\usepackage{dcolumn}
\usepackage{bm}
\usepackage{color}
\usepackage{soul}
\usepackage{ulem}
\usepackage{xfrac}

\begin{document}

\title{Symmetry Breaking in Photonic Crystals: On-Demand Dispersion from Flatband to Dirac Cones}

\author{H.~S.~Nguyen}
\email{hai-son.nguyen@ec-lyon.fr}
\author{F.~Dubois}
\author{T.~Deschamps}
\author{S.~Cueff}
\author{A.~Pardon}
\author{J-L.~Leclercq}
\author{C.~Seassal}
\author{X.~Letartre}
\author{P.~Viktorovitch}
\affiliation{Institut des Nanotechnologies de Lyon, INL/CNRS, Universit\'e de Lyon, 36 avenue Guy de Collongue, 69130 Ecully , France}

\date{\today}

\begin{abstract}
We demonstrate that symmetry breaking opens a new degree of freedom to tailor the energy-momentum dispersion in photonic crystals. Using a general theoretical framework in two illustrative practical structures, we show that breaking symmetry enables an on-demand tuning of the local density of states of a same photonic band from zero (Dirac cone dispersion) to infinity (flatband dispersion), as well as any constant density over an adjustable spectral range. As a proof-of-concept, we experimentally demonstrate the transformation of a very same photonic band from conventional quadratic shape to Dirac dispersion, flatband dispersion and multivaley one, by finely tuning the vertical symmetry breaking. Our results provide an unprecedented degree of freedom for optical dispersion engineering in planar integrated photonic devices.
\end{abstract}

\maketitle

Engineering the energy-momentum dispersion of photonic structures is at the heart of contemporary optics research. This fundamental feature molds the light propagation~\cite{Martinez1984}, dictates the coupling  with free space~\cite{Fan1997}, and tailors light-matter interactions~\cite{Noda2007}. Such a dispersion engineering is generally achieved through a designed periodic arrangement of materials with different permittivities in photonic crystals, metamaterials and metasurfaces. Recently, two particular types of dispersions have been intensively studied: flatband dispersion~\cite{Bergman2008,Sun2008,Jacqmin2015,Baboux2016,Vicencio2015,Mukherjee2015,Wang2015,Xu2015,Byrnes2016,Travkin2017} and Dirac dispersion~\cite{Huang2011,Sakoda2012,Moitra2013,Li2015,Treidel2010,Zhang2008,Abad2012,Chua2014,Haldane2008,Rechtsman2013,Rechtsman2013bis,Lu2015,Wang2009,Wu2015}. The first one provides slow light of zero group velocity with high density of states for a broad range of the Brillouin zone, thus greatly enhances light-matter interaction and nonlinear behaviors for low-threshold micro-lasers and information processing applications~\cite{Krauss2008,Soljacic2002}. Moreover, flatband gives rise to localized stationary eigenstates which are extremely sensitive to disorder effects due to an infinite effective mass~\cite{Baboux2016,Faggiani2016}. This suggests new regime of light localization~\cite{Vicencio2015,Mukherjee2015} other than conventional concepts such as Anderson localization~\cite{Schwartz2007,Segev2013} and optical bound states in continuum~\cite{Plotnik2011,Hsu2013}.  Being an opposite extreme to flat dispersion, Dirac dispersion (double Dirac cones with no bandgap) corresponds to massless photonic states. By analogy with the propagation of electrons in graphene, Dirac photons propagation could lead to phenomena such as Klein tunneling~\cite{Treidel2010} and \textit{Zitterbewegung}~\cite{Zhang2008} for photons.  Moreover, photonic Dirac dispersion opens the way to realize large-area single mode lasers~\cite{Abad2012,Chua2014}; and enables many exotic physical features such as zero-refractive index materials for transformation optics applications~\cite{Huang2011,Moitra2013,Li2015} and photonic topological insulator~\cite{Rechtsman2013,Rechtsman2013bis,Haldane2008,Lu2015,Wang2009,Wu2015}. Due to their completely opposite characteristics, flatband and Dirac dispersion are usually attributed to different bands of the photonic structures (2D tight-binding lattices~\cite{Sun2008,Jacqmin2015,Vicencio2015,Mukherjee2015}, accidental degeneracy in 2D photonic crystal~\cite{Huang2011,Sakoda2012,Moitra2013,Li2015}). Other configurations exhibit the sole presence of flatband states (1D tight-binding lattices~\cite{Baboux2016,Travkin2017}, dispersion engineering with hybrid micro cavities~\cite{Wang2015,Byrnes2016}). 

In this Letter, we propose a general and simple theoretical approach of High-index-Contrast subwavelength dielectric Gratings (HCGs) with broken symmetry, which are shown to provide a new degree of freedom for the design of photonic dispersion, and hence for the control of spatial and spectral characteristics of light. We show that breaking symmetry opens the way to the generation of any local density of photonic states from zero (Dirac cone) to infinity (flatband), as well as any constant density over an adjustable spectral range for the same photonic band. Such a unique concept is emphasized and exemplified with two illustrative cases: ``\textit{fishbone}" and ``\textit{comb}" structures. As a proof-of-concept, the transformation of a Dirac-band to a flatband in a ``\textit{comb}" grating structure is experimentally demonstrated by simply tuning the vertical symmetry breaking. We emphasize that our approach is very generic and can be applied to the design of a wide variety of devices for free space as well as planar integrated photonics.

\begin{figure}[t]
\begin{center}
\includegraphics[width=8cm]{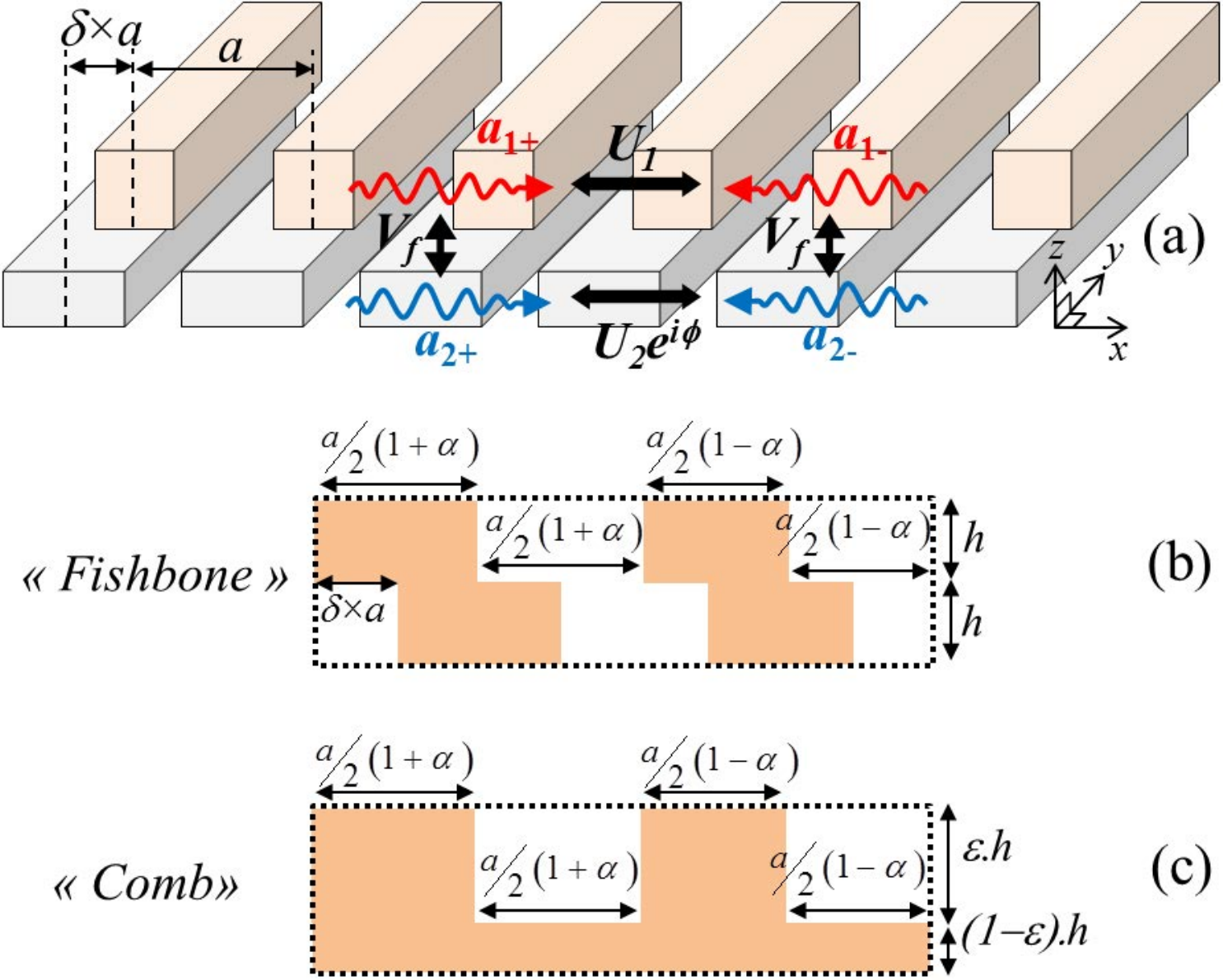}
\end{center} 
\caption{(Color online). (a) Sketch of a HCG with broken vertical symmetry. (b,c) Elementary cell of ``\textit{fishbone}"(b) and ``\textit{comb}"(c) structure with double period perturbation.}
\label{fig1}
\end{figure}

Inside the wide family of periodic photonic structures, HCGs have played a fast growing part during the last 20 years~\cite{Vikto2010,Letartre2003,Hasnain2012}.  Apart from rare exceptions~\cite{Suh2003,Shuai2013,Campione2016} reports in the literature are essentially dedicated to HCGs with non-broken vertical symmetry. HCG with broken vertical symmetry~\cite{Suh2003,Shuai2013,Campione2016} can be represented by two symmetric HCGs of same period $a$ in close near field proximity and with a lateral offset $\delta\times a$ as schematized in Fig.\ref{fig1}(a). In the following, we present an intuitive analytical model to describe such structures.  At this stage, the considered modes operate below the light cone. We will discuss later the situation where the guided modes lie partly inside the light cone and behave as Bloch resonances. In our model,  the $\omega(k_x)$  dispersion characteristic will be derived from different coupling processes undergone by the forward $(a_{1+},a_{2+})$ and backward $(a_{1-},a_{2-})$ fundamental zero-order waves of the two non-corrugated waveguide structures. The dispersion engineering is focused in the $(\omega,k_x)$ region in the vicinity of the first Brillouin zone boundary (i.e. $X$ point), where first-order diffractive coupling processes between backward and forward wave components are the most effective. A phenomenological description of coupling processes within the two symmetric gratings as well as of cross coupling processes occurring between them can be obtained using the coupled mode theory formalism. In the base formed by $[a_{1+},a_{1-},a_{2+},a_{2-}]$, the equations of coupled mode theory end up in a $4\times4$  Hermitian Hamiltonian $H$, given by: 
\begin{widetext}
\begin{equation}
	H=
	\begin{bmatrix}
      \omega_1+v_1k_x & U_1+\beta_1U_2e^{i\phi} & V_f	&  0 \\
			U_1+\beta_1U_2e^{-i\phi} & \omega_1-v_1k_x & 0 & V_f \\
			V_f & 0 & \omega_2+v_2k_x & U_2e^{i\phi}+\beta_2U_1 \\
			0 & V_f & U_2e^{-i\phi}+\beta_2U_1 & \omega_2-v_2k_x \\
	\end{bmatrix}			
\end{equation}
\end{widetext}
$U_{1,2}$ are the diffractive coupling rates between the forward and backward waves in each of the two grating structures when considered as far apart. The factors $\beta_{1,2}$ account for a supplementary diffractive coupling rate, which is induced by one grating on the evanescent part of the guided modes of the other grating. The coefficients $e^{\pm i\phi}$ with $\phi=2\pi\delta$ correspond to the first-order difrraction phase-shift at X point between the two gratings due to the lateral offset. $\omega_{1,2}$ and $v_{1,2}$ are respectively the energies and the group velocities of the propagating waves in the two non-corrugated waveguiding structures (the first Brillouin boundary is taken as the origin of the $k_x$ vector coordinates). $V_f$ stands for the evanescent coupling rate between waves of different gratings but propagating in the same direction. The anti-diagonal terms of $H$ correspond to the coupling processes between forward and backward waves of different gratings. These couplings are second-order terms that can be neglected in our model. Eigenfunctions or dispersion characteristics can be derived from diagonalization of $H$, yielding the eigenmode characteristics. Intuitively, breaking the vertical symmetry provides a very versatile tool for dispersion engineering of the coupled grating system: even and odd modes of the structure are no more eigenstates since they are allowed to couple and their hybridization results in a wide range of eigenmodes endowed with a rich variety of dispersion characteristics. Such hybridization requires a spectral overlap of even and odd modes, which is achieved when the diffractive coupling rates $U_1$,$U_2$ exceed the evanescent coupling rate $V_f$. This requirement is typically satisfied in high index contrast structures. With this approach, we derive a physically insightful analytical model of dispersion characteristics, which we confront below to RCWA (Rigorous Coupled-Wave Analysis) simulations in selected illustrating examples.

To illustrate the dispersion engineering concept presented previously, two cases studies  are investigated [see Fig.\ref{fig1}(b),(c)]. The first structure, called ``\textit{fishbone}",  is formed by two super-imposed identical gratings with adjustable lateral alignment [Fig.\ref{fig1}(b)]. This simple structure can be used to continuously explore  a wide range of vertical symmetry configurations, starting from the classical case of non broken vertical symmetry, where gratings are aligned or in phase (off-set ratio $\delta=0$, no phase-shift), all the way to the case of misaligned gratings by half a period ($\delta=0.5$, or $\phi=\pi$). In this example, we will consider a ``\textit{fish-bone}" structure consisting of two gratings of hydrogenated amorphous silicon (a-Si) with refractive index $n_{aSi}=3.15$, thickness $h=0.21\,\mu m$, period $a=0.35\,\mu m$, and surrounded by silica of refractive index $n_{SiO_2}$=1.5. The offset $\delta$ is the only joystick needed to ``shape" the dispersion.  

\begin{figure}[t]
\begin{center}
\includegraphics[width=8.5cm]{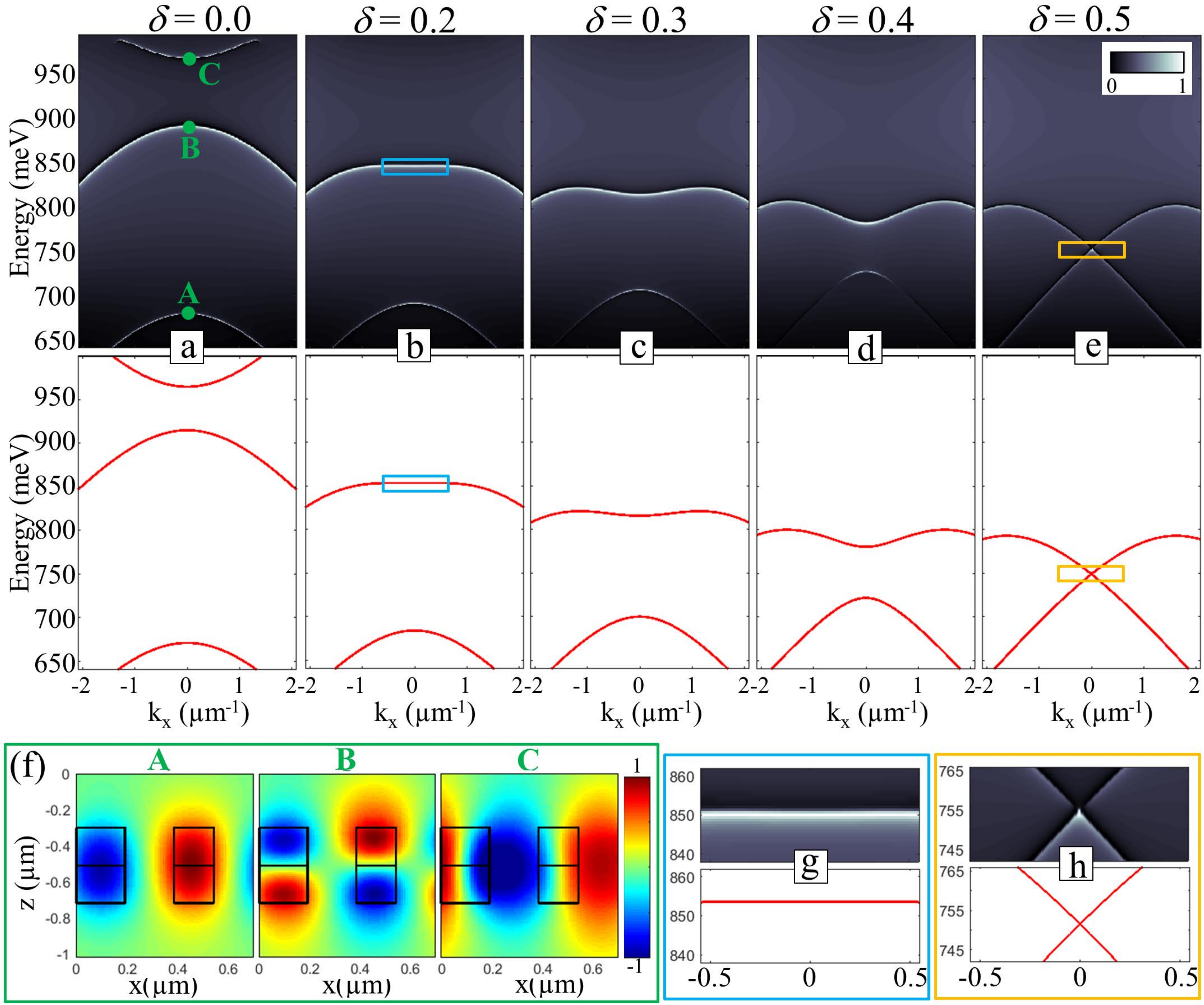}
\end{center}
\caption{(Color online). (a-e) Upper-part: Numerical simulation of the angle-resolved reflectivity spectrum with TE-polarized incident light for different values of $\delta$ of the ``\textit{fishbone}" structure; Lower-part: Analytical calculation of the photonic dispersion using the Hamiltonian based theoretical modeling. (f) Numerical simulation of the electric-field distribution of the three modes shown in (a). (g,h) Zoom of the rectangular selection area of (b,e) to highlight the flatband and Dirac cones.}
\label{fig2}
\end{figure}

We first investigate the reflectivity/transmission characteristics of the structure with the help of RCWA analysis. While such method cannot give direct access to wave-guided optical modes below the light cone, the latter can however be brought from the boundary ($X$ point) to the center ($\Gamma$ point) of the first Brillouin zone by a double-period design perturbation approach with perturbation magnitude $\alpha$ [see Fig.\ref{fig1}(b)]~\cite{Tran2010,Milord2015}. Wave-guided modes are then turned into wave-guided resonances, whose radiative lifetime $\tau_{rad}$ depends only on $\alpha$ ($\tau_{rad}\propto\alpha^{-2}$). Our theoretical model with the Hamiltonian $H$ is still valid if the radiative coupling is much smaller than the other couplings (i.e. diffraction and evanescent), thus requiring $\alpha\ll 1$. From now on, we choose $\alpha=0.1$. The upper part of Figs.\ref{fig2}(a-e) depicts RCWA calculations of the angularly resolved reflectivity spectrum with TE-polarized incident light (i.e. along $y$-axis) for different values of $\delta$. At $\delta=0$ [aligned gratings, see Fig.\ref{fig2}(a)], one may readily identify the two fundamental wave-guided modal components, which split into the high index dielectric (a-Si) and low index dielectric (silica) modes, separated by the first bandgap. One may also observe the first excited a-Si mode. As expected, the field distribution [Fig.\ref{fig2}(f)] shows a vertical even (odd) symmetry for the fundamental ($1^{st}$ excited) modes; and light is primarily concentrated in a-Si (silica) for high (low) refractive index modes. Misaligning the two gratings ($\delta\neq 0$) breaks the vertical symmetry and results in profound changes in the dispersion characteristics, as a result of strong coupling processes and of hybridization occurring between even and odd modes. At $\delta=0.2$, [see Figs.\ref{fig2}(b),(g)], a flatband  is generated at $\Gamma$, where not only the group velocity but also the curvature (second derivative) of the dispersion characteristic vanishes. Note that the phase-shift $\phi=2\pi\delta=0.4\pi$ corresponds to $cos\phi=0.31$, which is very close to the optimum value $1/3$ predicted by the analytical model, corresponding to a minimum of the fourth derivative~\cite{Supp}. Increasing $\delta$ further results in the formation of multivalley dispersions: multiple extremes separated by quasi-linear regions in the dispersion characteristic [see Figs.\ref{fig2}(c,d)], whose spectral range and slope appear to be finely controlled by the choice of $\delta$. Finally, with $\delta=0.5$, dispersion curves exhibit Dirac cones at $\Gamma$ [see Figs.\ref{fig2}(e,h)]. It is important to highlight that this degeneracy is not ``accidental" and is robust to variation of $h$ and $a$; the only condition being that the diffractive couplings of the two sub-gratings exhibit a $\pi$ phase-shift between them (i.e. $\delta=0.5$)\cite{Supp}.

The lower-part of Figs.\ref{fig2}(a-e) depicts analytical calculation of the photonic dispersion using the Hamiltonian based theoretical model described previously. Note that only 4 fitting parameters are required (the two gratings are identical), and a single set of parameters is used to fit the results of RCWA simulations for every value of $\delta$: $\omega_{1,2}=940\,meV$, $U_{1,2}=147\,meV$, $v_{1,2}=80\,meV.\mu m$, $V_f=119\,meV$. The factors $\beta_{1,2}$ are chosen to be zero since the supplemental diffractive coupling is negligible with respect to the direct diffractive coupling. The analytical calculations reproduce perfectly all dispersion characteristics given by the numerical simulations, in particular the flatband and Dirac dispersion. Thus our analytical model is successfully validated. 
\begin{figure}[t]
\begin{center}
\includegraphics[width=8.5cm]{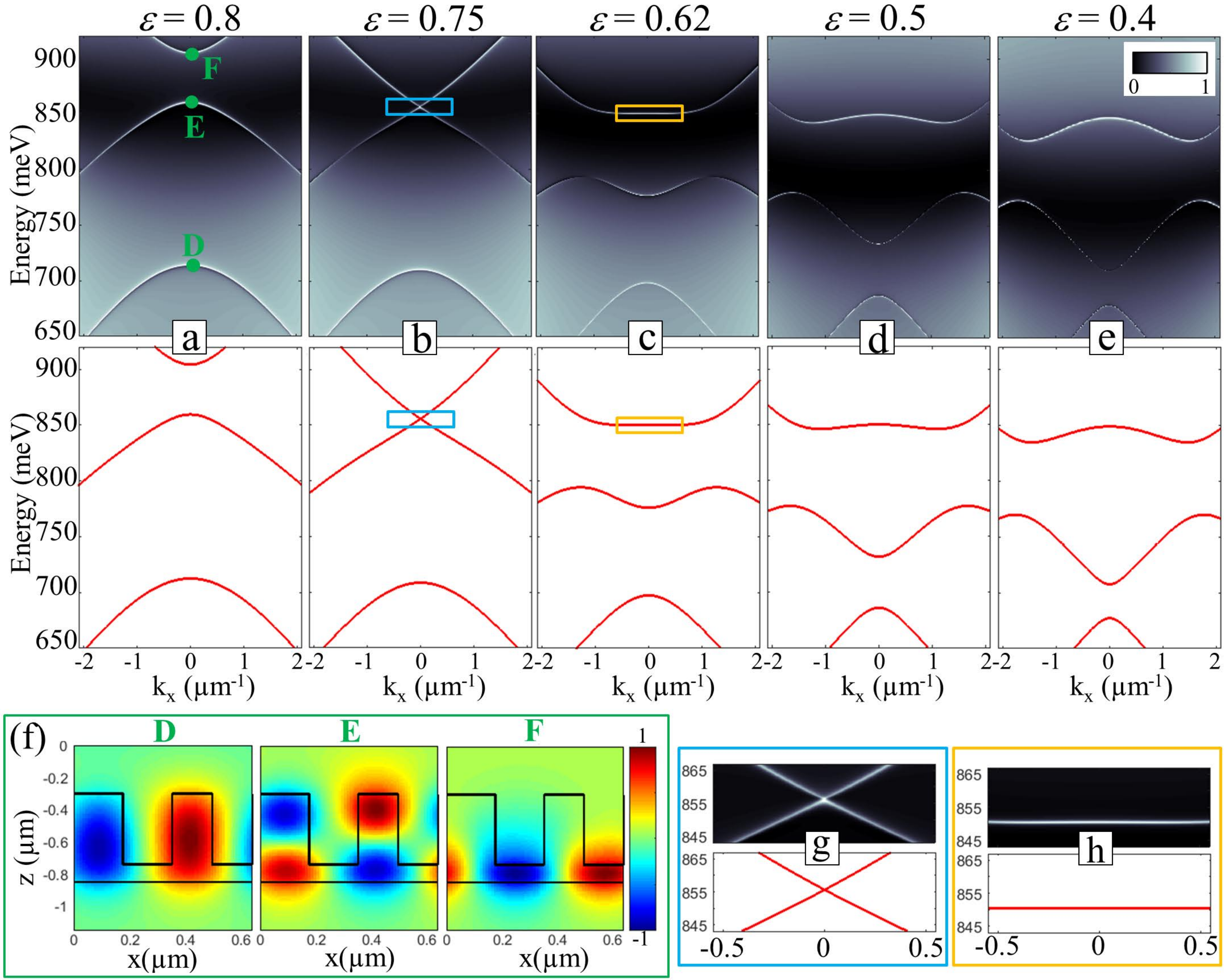} 
\end{center}
\caption{(Color online). (a-e) Upper-part: Numerical simulation of the angle-resolved reflectivity spectrum with TE-polarized incident light for different values of $\epsilon$ of the ``\textit{comb}" structure; Lower-part: Analytical calculation of the photonic dispersion using the Hamiltonian based theoretical modeling. (f) Numerical simulation of the electric-field distribution of the three modes shown in (a). (g,h) Zoom of the rectangular selection area of (b,e) to highlight the flatband and Dirac cones.}
\label{fig3}
\end{figure}

The second structure, called \textit{``comb"} structure [see Figs.\ref{fig1}(c)] is representative of a more standard configuration. It consists of an asymmetric a-Si grating, which is viewed as two superimposed non separated symmetric gratings, the second ``grating" being non corrugated. A straightforward ``joy-stick" to tune the vertical symmetry breaking is the etch depth ratio $\epsilon$ (etch depth $= \epsilon\times h$, where $h$ is the total thickness of the grating), which may span the range 0 to 1. In this example, we choose the thickness $h=0.55\,\mu m$ and the period $a=0.32\,\mu m$. Figures \ref{fig3}(a-e) show RCWA simulations of dispersion characteristics versus analytical calculations, for different $\epsilon$. Increasing the asymmetry results in a dramatic modulation of dispersion curves:  quadratic dispersions at low symmetry breaking [$\epsilon$ = 0.8, see Fig.\ref{fig3}(a)] transform to Dirac cones [$\epsilon$ = 0.75, see Figs.\ref{fig3}(b,g)], flatband [$\epsilon$ = 0.62, see Figs.\ref{fig3}(c,h)], then \textit{M-shaped} and \textit{W-shaped} multivalley dispersions [Figs.\ref{fig3}(d,e)]. Again, the dispersion engineering is achieved via the hybridization between odd and even guided modes carried in the structures, shown in Fig.\ref{fig3}(f). But differently to the \textit{``fishbone"} case, the Dirac dispersion of \textit{``comb"} structure  corresponds to an accidental degeneracy~\cite{Huang2011,Sakoda2012} between an odd and an even mode of the structure. In other words, the value of $\epsilon$ corresponding to Dirac dispersion depends strongly on the value of the ratio $a/h$. 
\begin{figure}[t]
\begin{center}
\includegraphics[width=8.5cm]{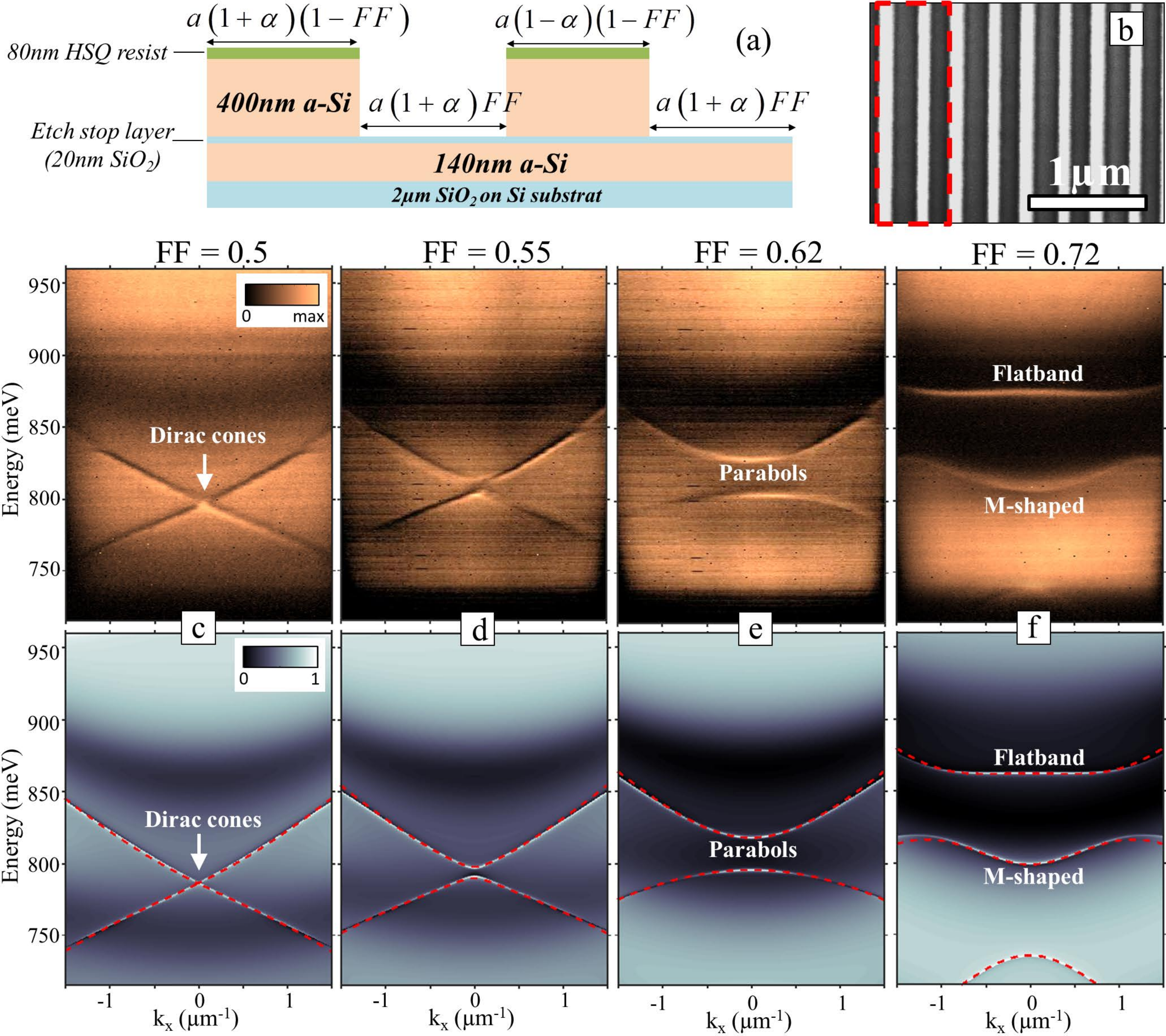} 
\end{center}
\caption{(Color online). (a) Sketch of the sample design. (b) SEM images (top-view) correspond to structure of $FF=0.72$. The scale bar is $1\,\mu m$. Rectangular selection indicate an elementary cell. (c-f) Upper part: Experimental angle-resolved reflectivity spectrum with TE-polarized incident light for different values of $FF$; Lower part: Numerical simulation of the angle-resolved reflectivity spectrum with TE-polarized incident light for different values of $FF$. Red dashed lines are analytical calculations using the Hamiltonian based model.}
\label{fig4}
\end{figure}

As a proof-of-concept, we experimentally demonstrate  Dirac cones, flatband and multivalley dispersions in \textit{``comb"} structure. Our sample consists of asymmetric a-Si gratings on silica, fabricated via electron beam lithography and ionic dry etching~\cite{Supp}. Note that a transformation from Dirac cones to flatband requires only a slight modification of the vertical symmetry breaking [see Figs.\ref{fig3}(b,c)]. It allows to adopt a practical approach to obtain different dispersion characteristics for the same layer stack and etching depth: i) All structures ($80\mu m\times80\mu m$) have the same etch depth ratio $\epsilon=0.74$, thus exhibit strong vertical symmetry breaking; ii) The odd-even mode coupling is finely tuned by varying the filling factor $FF$ of the top grating. The sample design is presented in Fig.\ref{fig4}(a). The top grating is of period $a=0.31\,\mu m$, $0.4\,\mu m$ thickness and perturbation magnitude $\alpha=0.1$. The thickness of the non-corrugated waveguide is $0.14\,\mu m$ . Scanning Electron Microscope (SEM) image of the structure corresponding to $FF=0.72$ is shown in Figs.\ref{fig4}(b). A white optical beam (halogen lamp) is focused onto the sample via a microscope objective (NA=0.42). The reflectivity is collected via the same objective and then captured in Fourier space onto the sensor of an infrared camera; the latter is coupled to a spectrometer~\cite{Supp}. The upper panels of Figs.\ref{fig4}(c-f) show experimental angle-resolved reflectivity spectrum of TE polarization when the filling factor is varied from $FF=0.5$ to $FF=0.72$. A transformation from Dirac dispersions [Fig.\ref{fig4}(c)] to flatband [Fig.\ref{fig4}(f)] of the same photonic band is clearly evidenced. We also highlight the observation of a \textit{M-shaped} dispersion [Fig.\ref{fig4}(f)]. The experimental results are perfectly consistent with both the RCWA numerical simulations and the analytical calculations [see lower panels of Figs.\ref{fig4}(c-f)]. Note that the refractive index provided by ellipsometry measurements are used as input for RCWA simulations. The slight energy shift between the observed spectra and the theoretical prediction is likely due to thickness and groove size difference. 

While dispersion engineering by mixing modes of different natures has been suggested by different groups~\cite{Wang2015,Byrnes2016}), our theoretical proposal and experimental demonstrations show an unprecedented degree of freedom to tailor the dispersion characteristics of photonic structures, whose curvature can be finely tuned from zero to infinity. This opens an unique playground for both exotic Dirac and flatband physics. In particular, it will be possible to study the light delocalization when a dispersion is gradually transformed from flatband (exceptional localized states and ultra sensitive to disorders~\cite{Baboux2016,Faggiani2016,Vicencio2015,Mukherjee2015}) to Dirac cones (very robust versus disorder effect of Anderson localization~\cite{Deng2015}). 

Another attractive prospect is the integration of active material in these designs, for example by placing quantum wells in the non corrugated part of the ``\textit{comb}" structure, or by depositing monolayers of 2D material on top of the structure via exfoliation. In the weak light-matter coupling regime, one can expect to obtain ultra-compact, low-threshold micro-lasers with flatband states; and large area single mode with Dirac states~\cite{Abad2012,Chua2014}. In addition, linear dispersion characteristics with on-demand group velocity can also be obtained: the group velocity around the Dirac point can be finely controlled from the group velocity of the non-corrugated waveguides down to theoretically zero~\cite{Supp}. These remarkable features are highly desired in a variety of non-linear photonics applications such as ultra-compact mode-locked lasers. Moreover, our concept can be used in strong coupling regime~\cite{Weisbuch1992} to control the properties of polariton condensates and lasers~\cite{Kasprazak2006,RMPIacopo2013}. Note that dispersion engineering in polaritonic systems has been mostly achieved by patterning Fabry-Perot cavity into microstructure, which requires etching the active layer~\cite{Wertz2010,Nguyen}. Adopting our design, the active layer will be unpatterned - an ideal configuration to obtain the strong coupling regime with fragile 2D materials~\cite{Liu2015}. We also highlight the possibility to use multivalley dispersion in the strong coupling regime to obtain spontaneous momentum symmetry breaking and two-mode squeezing~\cite{Sun2017}, as well as the observation of the Josephson effect in momentum space~\cite{Troncoso2014}. 

In conclusion, we have demonstrated both theoretically and experimentally how to utilize the vertical symmetry breaking to engineer the dispersion characteristics of photonic structures. For the first time, a very same dispersion band can be finely transformed from conventional quadratic shape to Dirac cones, flatband and multivalley one. We would like to emphasize that the theoretical approach developed in the present work is very generic in that it offers general design rules and predicts the universal set of dispersion characteristics for a wide range of configurations. Such generality applies to any type of photonic structure involving two coupled gratings with broken symmetry. This includes HCG membrane devices with vertical broken symmetry operating above the light cone as well as HCG waveguide devices with lateral broken symmetry operating below the light cone. Plenty of room is left for an extra wide variety of configurations, with, among other prospects, the extension to 2D photonic crystal slab, which naturally lend themselves to additional degrees of freedom in 3D manipulation of light (e.g. angular and polarization resolution).  

The authors gratefully acknowledge R. Orobtchouk for the in-house developed RCWA codes, and T. C. H. Liew for fruitful discussions. The authors would like to thank the staff from the NanoLyon Technical Platform for helping and supporting in the realization of the structures. This work is partly supported by the French National Research Agency (ANR) under the project PICSEL (ANR-15-CE24-0026).

\end{document}